\documentclass[aps,pra,twocolumn,showpacs,floatfix]{revtex4}

\usepackage{epsfig}
\usepackage{graphicx}
\usepackage{dcolumn}
\begin{document}
\title{Frequency-dependent polarizabilities of alkali atoms from ultraviolet
through infrared spectral regions}
\author{M. S. Safronova}
\affiliation{Department of Physics and Astronomy, University of Delaware, Newark, Delaware 19716}
\author{Bindiya Arora}
\affiliation{Department of Physics and Astronomy, University of Delaware, Newark, Delaware 19716}
\author{Charles W. Clark}
 \affiliation{
Physics Laboratory,
National Institute of Standards and Technology,
Technology Administration,
U.S. Department of Commerce, 
Gaithersburg, Maryland 20899-8410}
\date{\today} 
\begin{abstract}
We present results of first-principles calculations of the
frequency-dependent polarizabilities of all alkali atoms for light in the
wavelength range 300-1600 nm, with particular attention to wavelengths of
common infrared lasers.  We parameterize our results so that they can be
extended accurately to arbitrary wavelengths above 800 nm.  This work is
motivated by recent experiments involving simultaneous optical trapping of
two different alkali species.  Our data can be used to predict the
oscillation frequencies of optically-trapped atoms, and particularly the
ratios of frequencies of different species held in the same trap.  We
identify wavelengths at which two different alkali atoms have the same
oscillation frequency.  
\end{abstract}

\maketitle

\section{Introduction}

The trapping of two different species of ultracold gases has been an active
topic of research for about ten years (see \cite{Lundblad} and references therein), and has led to the
simultaneous production of quantum degenerate Bose-Einstein and Fermi-Dirac
gases \cite{schreck:080403,hadzibabic:160401,Modugno} and recently to the observation of heteronuclear Feshbach resonances involving Bose-Einstein and Fermi-Dirac species \cite{Stan,Inouye}.

Two recent experiments \cite{Stan,Inouye} provide a glimpse of the possibilities of
mapping out the phase diagram of a Bose-Einstein/Fermi-Dirac mixture,
potentially a rich system that remains to be explored.  Our present work
provides accurate estimates of the trapping frequencies of disparate
species confined in the same optical trap.  We have been attentive to finding
cases in which all species, Bose-Einstein and Fermi-Dirac
atoms, and the molecules formed from them, have the same trapping
frequencies. This case is of particular interest because it has been shown to be optimal for a recently proposed scheme of cooling mediated by 
Feshbach resonances \cite{Miguel}.  In
addition, we note that while the DC polarizabilities of alkali atoms have
been subject to careful experimental investigation, there is relatively
little accurate experimental data on frequency-dependent polarizabilities.
We draw attention to the fact that current experiments offer the possibility
of measurements of unprecedented accuracy regarding the ratio of
frequency-dependent polarizabilities of different species.  Such
measurements could provide more decisive comparison of {\it ab initio}
atomic theory with experiments, than has been made to date.

We begin with a summary of the calculation of frequency-dependent
polarizabilities, the details of which have been presented in previous
publications \cite{gate,rb}.  The computational method we use here has delivered
agreement with experiments on static polarizabilities at the 1\% level.
There are few direct measurements of frequency-dependent polarizabilities -
the experiments are much more demanding, and have larger uncertainties
associated with modeling the intensity profile of a focused laser beam.
However, the alkali metal atoms are a special case, at least in the infrared spectral
region.  There, the polarizability is dominated by the contribution from the
longest-wavelength resonance transitions, and it is possible to perform
consistency checks between static values and long-wavelength values of the
frequency-dependent atomic polarizabilities.  We present tabulated values of
calculated frequency-dependent polarizabilities over a range of wavelengths
from the ultraviolet through the infrared spectrum, with particular
attention to those wavelengths (primarily in the infrared) which have been
or might be employed in the present generation of optical traps.  For the
infrared region, we display simple formulae that allow accurate
computation of frequency-dependent polarizabilities for wavelengths not
explicitly reported upon here.

Finally, we enumerate selected wavelengths, at which two species of
optically-trapped alkali atoms would have the same frequencies of
oscillation in a common optical trap.  Such wavelengths offer opportunities
for mapping out phase diagrams of Bose-Einstein and Fermi-Dirac mixtures in
a space of reduced dimensionality.  Moreover, to the extent that molecular
resonances consist of weakly bound states at large internuclear separation,
the oscillation frequencies of trapped molecular species should be very
close to the frequencies of their atomic components.  This may present an
ideal system for investigation. 
\section{Calculation of polarizabilities}
 \begin{table*} [ht]
\caption{\label{tab1} The values of the frequency-dependent polarizabilities of alkali-metal atoms in $a^3_0$. 
The wavelengths (in air) are given in nm, the corresponding frequencies are given in a.u.}
 \begin{ruledtabular}
\begin{tabular}{rdrrrrrr}
\multicolumn{1}{c}{$\lambda_{air}$} &
\multicolumn{1}{c}{$\omega$} &
\multicolumn{1}{c}{Li} & 
\multicolumn{1}{c}{Na} & 
\multicolumn{1}{c}{K} &
\multicolumn{1}{c}{Rb} & 
\multicolumn{1}{c}{Cs}&
 \multicolumn{1}{c}{Fr} \\
    \hline
 1554 &  0.02931 & 201.0(7) & 189.4(2) &  381.2(8)  &   424.0(7) &   570.9(1.2) &   407.9(2.0)\\
 1550 &  0.02939 & 201.3(7) & 189.6(2) &  381.8(8)  &   424.7(7) &   572.2(1.2) &   408.5(2.0)\\
 1540 &  0.02958 & 201.9(7) & 190.0(2) &  383.4(8)  &   426.6(7) &   575.6(1.2) &   410.1(2.0)\\
 1340 &  0.03399 & 218.1(8) & 201.0(2) &  428.2(9)  &   479.5(7) &   673.7(1.3) &   455.0(2.1)\\
 1332 &  0.03420 & 219.0(8) & 201.6(2) &  430.7(9)  &   482.5(7) &   679.5(1.3) &   457.5(2.1)\\
 1313 &  0.03469 & 221.1(8) & 203.0(2) &  436.9(9)  &   489.9(7) &   694.1(1.3) &   463.7(2.1)\\
 1240 &  0.03673 & 231.0(8) & 209.3(2) &  465.9(9)  &   524.6(7) &   764.6(1.4) &   492.8(2.2)\\
 1152 &  0.03954 & 247.0(8) & 219.4(2) &  516.1(1.0)&   585.7(8) &   899.9(1.6) &   543.6(2.3)\\
 1090 &  0.04179 & 262.7(8) & 228.8(2) &  568.7(1.1)&   650.8(8) &  1062.3(1.7) &   597.2(2.4)\\
 1064 &  0.04281 & 270.8(8) & 233.6(2) &  597.5(1.2)&   686.9(9) &  1162.1(1.9) &   627(3)\\
 1060 &  0.04297 & 272.1(8) & 234.3(2) &  602.4(1.2)&   693.1(9) &  1180.0(1.9) &   632(3)\\
 1053 &  0.04326 & 274.6(8) & 235.7(3) &  611.3(1.2)&   704.4(9) &  1213.2(1.9) &   641(3)\\
 1047 &  0.04351 & 276.7(8) & 237.0(3) &  619.3(1.2)&   714.5(9) &  1243.8(2.0) &   650(3)\\
 1030 &  0.04422 & 283.2(8) & 240.7(3) &  643.9(1.3)&   746.0(9) &  1343.6(2.1) &   675(3)\\
  985 &  0.04624 & 304.0(8) & 252.0(3) &  728.3(1.4)&   855.8(1.0)&     1759(3) &   765(3)\\
  980 &  0.04648 & 306.7(8) & 253.5(3) &  739.9(1.4)&   871.2(1.0)&     1828(3) &   778(3)\\
  946 &  0.04815 & 327.8(9) & 264.4(3) &  836.4(1.6)&  1001.6(1.1)&     2581(4) &   886(3)\\
  930 &  0.04898 & 339.6(9) & 270.3(3) &  895.9(1.7)&  1084.4(1.2)&     3325(5) &   957(3)\\
 799.3&0.05699 & 549.7(1.1) & 353.8(4) &  3666(7)   &   13483(14) &     -2333(4) & -1474(11)\\ 
  700 & 0.06507 &   1983(3)    &   553.1(0.6) &    -1394(3)  & -1192.7(1.4)  & -707.9(1.4) & -3787(18)\\
  600 & 0.07592 &  -645.9(1.3) &  4506(5) &  -438.4(0.9) &  -421.8(0.7)  &  -333.9(1.0)&   -523(3)\\
  500 & 0.09110 &  -200.1(0.8) &  -411.1(0.4) &  -201.3(0.5) &  -196.8(0.6)  &  -166.3(1.0)&   -204(2)\\  
  400 & 0.11388 &   -86.4(0.8) &  -134.0(0.2) &  -131.4(2.4) &  -109.7(1.1)  &   -88.2(1.3)&   -111(4) \\ 
  300 & 0.15183 &   -38.4(1.1) &   -57.5(0.3) &   -47.2(0.4) &   -43.7(1.1)  &   -35(2)    &   -43(2) \\  
 \end{tabular}
\end{ruledtabular}
 \end{table*} 
The frequency-dependent polarizability of an atom in its ground state may be separated into 
polarizability of the ionic core $\alpha_{\rm core}$ and the  valence contribution $\alpha_v^{ns}$.
The calculation of the core polarizability is carried out in the random-phase approximation (RPA);
we verified that our static values agree with RPA values from Ref.~\cite{jkh}.
The separation of the polarizability into core and valence parts also requires the addition of the 
compensation term $\alpha_{vc}$ which accounts for the contribution from the excitation to the occupied 
valence shell that is forbidden by the Pauli exclusion principle. The calculation of the core polarizability 
does not exclude the excitation from the core to the valence shell and half of 
this contribution has to be subtracted. Thus, the polarizability contributions may be separated as 
\begin{equation}
\alpha^{ns}=\alpha_{\rm core}+\alpha_{vc}+\alpha_v^{ns}.
\end{equation}
We calculate the valence part of the ac polarizability of the ground state $ns$
for the alkali-metal atoms by computing the sum over states 
 \begin{eqnarray}
\alpha_v^{ns}&=&\frac{1}{3}\sum_{n^{\prime}}
\left(
\frac{\delta E_{n^{\prime}p_{1/2}}
\langle n^{\prime}p_{1/2}\|D\| ns \rangle^2 }
{\delta E_{n^{\prime}p_{1/2}}^2-\omega^2}\right.  \nonumber \\
&+&\left. \frac{\delta E_{n^{\prime}p_{3/2}}
\langle n^{\prime}p_{3/2}\|D\| ns \rangle^2 }
{\delta E_{n^{\prime}p_{3/2}}^2-\omega^2}
\right),
\label{eq1}
\end{eqnarray}
where $\langle np\|D\| ns \rangle$ is the reduced electric-dipole matrix element, 
$\delta E_{n^{\prime}p_{1/2}}= E_{n^{\prime}p_{1/2}}-E_{ns}$,
and $\delta E_{n^{\prime}p_{3/2}}= E_{n^{\prime}p_{3/2}}-E_{ns}$.  
 In this formula,  $\omega$ is 
assumed to be at least several linewidths off resonance with the 
corresponding transition. We use the system of atomic units, a.u., in which 
$e/\sqrt{4\pi\epsilon_0}, m_{\rm e}$, and the reduced Planck constant $\hbar$ have the 
numerical value 1, in Eq.(\ref{eq1}).  Polarizability in a.u. has the dimensions of 
volume, and its numerical values presented here are thus measured 
in units of $a^3_0$, where $a_0\approx0.052918$~nm is the Bohr radius.
The atomic units for $\alpha$ can be be converted to SI units via
 $\alpha/h$~[Hz/(V/m)$^2$]=2.48832$\times10^{-8}\alpha$~[a.u.], where
 the conversion coefficient is $4\pi \epsilon_0 a^3_0/h$ and 
 Planck constant $h$ is factored out. The atomic unit of frequency $\omega$
 is $E_h/\hbar\approx4.1341\times10^{16}$~Hz, where $E_h$ is the Hartree
 energy.

  \begin{table*} [ht]
\caption{\label{tab2} The parameters $A^{ns}$ (in a.u.) for the calculation of ground state frequency-dependent polarizabilities of
 alkali-metal atoms  in $a^3_0$ for the wavelengths above 800~nm. The corresponding uncertainties in the values of $A^{ns}$ are given in the last 
column. The corresponding reduced matrix elements and energy differences are given in columns 
labeled $\langle np\|D\| ns \rangle$ and  $\delta E_{np}$.}
\begin{ruledtabular}
\begin{tabular}{ccccrrrr}
\multicolumn{1}{c}{Atom} &
\multicolumn{1}{c}{n} &
\multicolumn{1}{c}{$\left\langle np_{1/2} \left\|D\right\| ns\right\rangle$}&
\multicolumn{1}{c}{$\left\langle np_{3/2} \left\|D\right\| ns\right\rangle$}&
\multicolumn{1}{c}{ $\delta E_{np_{1/2}}$}&
\multicolumn{1}{c}{ $\delta E_{np_{3/2}}$}&
\multicolumn{1}{c}{$A^{ns}$} &
\multicolumn{1}{c}{$\delta A^{ns}$} \\
\hline
 Li & 2 & 3.317(4)   & 4.689(5)   &  0.067906 &  0.067907 & 2.04  & 0.69 \\
 Na & 3 & 3.5246(23) & 4.9838(34) &  0.077258 &  0.077336 & 1.86  & 0.12 \\
 K  & 4 & 4.102(5)   & 5.800(8)   &  0.059165 &  0.059428 & 6.26  & 0.33 \\
 Rb & 5 & 4.231(3)   & 5.977(4)   &  0.057314 &  0.058396 & 10.54 & 0.60 \\
 Cs & 6 & 4.4890(65) & 6.3238(73) &  0.050932 &  0.053456 & 17.35 & 1.00 \\
 Fr & 7 & 4.277(8)   & 5.898(15)  &  0.055758 &  0.063442 & 24.8  & 1.8 
\end{tabular}
\end{ruledtabular}
\end{table*}

The sum over $n^{\prime}$ in Eq.(\ref{eq1}) converges rapidly, and
unless the frequency is resonant with the transitions other than the
primary ones ($ns-np_{1/2}$ and $ns-np_{3/2}$), the first term is  dominant. 
Therefore, only a few terms need to be calculated accurately. 
We use the experimental numbers compiled in Ref.~\cite{relsd} together with their
uncertainties  for the first term, for example for $n^{\prime}=3$ term
for Na. High-precision theoretical all-order values are used for the next 
three terms (for example $n^{\prime}=4, 5, 6$ terms for Na). We refer the reader
to Refs.~\cite{cs,na,relsd} for the detailed description of the single-double (SD) all-order method
and its extensions.   
The all-order values of the matrix elements are evaluated for their accuracy, and the extensions of the SD all-order 
method (SDpT, which partially includes triple excitations or SD$_{sc}$, which includes semi-empirical scaling of dominant terms) are used for certain 
transitions where those values are expected to be of better accuracy. The uncertainty 
of the resulting matrix elements is evaluated  based on the relative
importance of certain classes of the all-order terms, extensive  
comparison of various atomic properties of alkali-metal atoms with experiment \cite{na,relsd,rb}, and the  
 spread of SD, SDpT,  and SD$_{sc}$ values.
 Some of these values were published previously 
in Refs.~\cite{relsd,csus}. The experimental values from \cite{csus}  are used for $6s-7p_{1/2}$
and $6s-7p_{3/2}$ transitions. In summary, the set of values used for the calculation of the 
first four terms in the sum over states  of Eq.(\ref{eq1}) consists of  the best known values for these transitions.
The experimental energies from Refs.~\cite{NIST,NIST1} are used for the corresponding energy
levels. We refer to the total contribution from the first four terms as the main term; the remaining contributions 
are referred to below as the tail contribution; i.e.
\begin{equation}
\alpha_v^{ns}=\alpha_{\rm main}(n^{\prime}=n,\dots, n+3)+\alpha_{\rm tail}(n^{\prime}>n+3).
\end{equation}

  \begin{table} 
\caption{\label{tab3} Selected wavelengths $\lambda_{\rm air}$ in nm, at which two species of
optically-trapped alkali atoms ($^6$Li, $^7$Li, $^{23}$Na, $^{40}$K, $^{41}$K, $^{87}$Rb, and $^{133}$Cs)
 would have the same frequencies of
oscillation in a common optical trap. The corresponding ground state frequency-dependent polarizabilities of the alkali-metal atoms (in a.u.) are 
also given.}
 \begin{ruledtabular}
 
\begin{tabular}{rcrcrc}
\multicolumn{1}{c}{$\lambda_{\rm air}$} &
\multicolumn{1}{c}{$1$} &
\multicolumn{1}{c}{$\alpha_1$} &
\multicolumn{1}{c}{$2$} &
\multicolumn{1}{c}{$\alpha_2$} \\
\hline
   557.56(13)  & $^6$Li &  -360(1)    &   $^{23}$Na&  -1375(1)    \\[0.2pc]
   768.648(3)  & $^6$Li &   681.3(1.3)  &   $^{40}$K &   4525(112)  \\
   799.46(12)  & $^6$Li &   549(1)    &  $^{40}$K &   3649(7)    \\[0.2pc]
   768.645(3)  & $^6$Li &   681.3(1.3)  &   $^{41}$K &   4638(112)  \\
   798.44(12)  & $^6$Li &   553(1)    &  $^{41}$K &   3763(7)    \\[0.2pc]
   786.07(1)  & $^6$Li &   597.9(1.2)&   $^{87}$Rb&   8639(20)   \\
   804.57(4)  & $^6$Li &   533(1)    &   $^{87}$Rb&   7704(8)    \\[0.2pc]
   862.097(28)  & $^6$Li &   412.5(1.0) & $^{133}$Cs &   9113(22)   \\
   904.76(4)    & $^6$Li &   361.7(9)  & $^{133}$Cs &   7991(12)   \\[0.2pc]
   549.72(18) & $^7$Li & -329(1) & $^{23}$Na & -1078(1) \\[0.2pc]
   768.665(3) & $^7$Li & 681(1) & $^{40}$K & 3880(112)\\
   806.9(2) & $^7$Li & 526(1) & $^{40}$K & 2998(6) \\ [0.2pc]
   768.662(3) & $^7$Li & 681(1) & $^{41}$K  & 3978(112) \\
   805.55(15) & $^7$Li & 530(1) & $^{41}$K  & 3096(6) \\[0.2pc]
   786.56(1) & $^7$Li & 595.9(1.2) & $^{87}$Rb & 7383(19)\\
   807.16(4) & $^7$Li & 526(1) & $^{87}$Rb & 6510(6)\\[0.2pc]
   863.436(32) & $^7$Li & 410.5(1.0) & $^{133}$Cs & 7777(20) \\
   907.25(6)  & $^7$Li & 359.3(9) & $^{133}$Cs & 6805(10) \\[0.2pc]
   789.240(7) & $^{23}$Na &   364.9(4)  &   $^{87}$Rb&   1380(16)   \\
   946.664(5) &  $^{23}$Na &   264.1(3)  &   $^{87}$Rb&    999(1)    \\[0.2pc]
   875.53(3)  &  $^{23}$Na &   295.5(3)  &   $^{133}$Cs&   1709(11)   \\
   1021.5(4)  &  $^{23}$Na &   242.6(3)  &   $^{133}$Cs&   1403(2)    \\[0.2pc]
   768.848(3) &  $^{40}$K  & -3893(116)  &   $^{87}$Rb&  -8470(10)   \\
   784.32(1)  &  $^{40}$K  &  6806(13)   &   $^{87}$Rb&  14800(27)   \\
   807.31(6)  & $^{40}$K  &  2968(6)    &   $^{87}$Rb&   6454(6)    \\[0.2pc]
   868.90(4)  &  $^{40}$K  &  1298(3)    &   $^{133}$Cs&   4317(14)   \\
   938.6(2)  &  $^{40}$K  &   862(2)    &   $^{133}$Cs&   2868(4)    \\[0.2pc]
   768.851(3) & $^{41}$K & -4013(116) & $^{87}$Rb & -8472(10) \\
   784.42(1) & $^{41}$K & 6767(12) & $^{87}$Rb & 14358(26) \\
   808.07(7) & $^{41}$K & 2916(6) & $^{87}$Rb & 6187(6) \\[0.2pc]
   869.14(4) & $^{41}$K & 1296(2)& $^{133}$Cs & 4203(14) \\
   940.9(2) & $^{41}$K & 854(2) & $^{133}$Cs & 2771(4)\\[0.2pc]
    790.303(6)  &  $^{87}$Rb & -1299(17)   &  $^{133}$Cs &  -1986(3)    \\
   873.39(4)  &  $^{87}$Rb &  1624(2)    &   $^{133}$Cs&   2483(12)   \\
   1158(3)   &  $^{87}$Rb &   580.5(8)  &   $^{133}$Cs&    888(2)    \\[0.2pc]
 \end{tabular}
\end{ruledtabular}
 \end{table}

We evaluated the tail contributions 
 with a complete set of basis Dirac-Hartree-Fock (DHF) wave 
functions generated using the B-spline method  \cite{Bspline}.
We use 40 splines of order $k=7$ for each angular momentum. 
The basis set orbitals are defined on a non-linear
  grid and are constrained to a large spherical cavity
  of a radius $R=75-110$~a.u.  The cavity radius is chosen to accommodate first four $ns$, $np_{1/2}$, 
  and $np_{3/2}$ valence orbitals. For consistency, we use 
  the same basis sets for the tail calculations as we used in the all-order 
  calculation of the main term matrix elements.
The tail contribution is calculated in the DHF approximation.
We list our results for the frequency-dependent polarizabilities 
from the ultraviolet through the infrared spectrum in Table~\ref{tab1}. Particular attention is given to the 
values for the infrared wavelengths  which have been
or might be employed in the present generation of optical traps.
 Our values of the ac polarizabilities are accurate to better than 0.5\% with the exception of the 
 values corresponding to wavelengths in the ultraviolet  spectrum. For these wavelengths, non-primary
 resonances become dominant and the uncertainty of the polarizability values is dominated by the 
 uncertainty in the corresponding resonance matrix elements. For example, the first resonances in Rb 
 ($5s-5p$ transitions) occur at  795~nm and 780~nm, while next resonances ($5s-6p$) transitions are near
  420~nm. As the accuracy of the $5s-6p$ matrix elements is lower than that of the $5s-5p$ 
  matrix elements the resulting accuracy of the ac polarizabilities at 400~nm is also lower.

 \begin{figure} 
 \hbox{\centerline{\epsfxsize=3.3in\epsffile{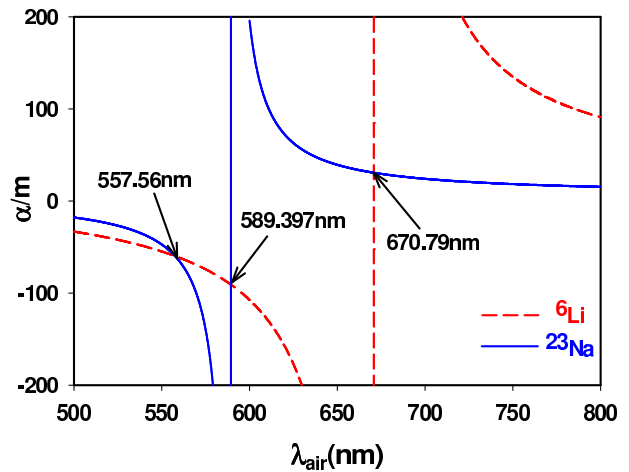}}}
    \caption{The ratios $\alpha(\lambda)/m$ of frequency-dependent 
	polarizabilities to the atomic weights for $^6$Li and $^{23}$Na.}   
\label{fig1}     
\end{figure}      

 \section{Results}
 
We have investigated the dependence on the frequency $\omega$ of all of the terms contributing to $\alpha^{ns}$, where $ns$ is the 
ground state. We find that the $\alpha_{core}$, $\alpha_{vc}$, $\alpha_{tail}$, and 
$\alpha^{ns}_v$ terms with $n^{\prime} = n+1, n+2, n+3$ depend weakly on the value of $\omega$
for wavelengths above 800~nm. In fact, it is possible to accurately parameterize the 
values of the ac polarizabilities for those wavelengths by the formula 
 \begin{eqnarray}
\alpha_v^{ns}(\omega)&=&\frac{1}{3}
\frac{\delta E_{np_{1/2}}
\langle np_{1/2}\|D\| ns \rangle^2 }
{\delta E_{np_{1/2}}^2-\omega^2}  \nonumber \\
&+&\frac{1}{3}\frac{\delta E_{np_{3/2}}
\langle np_{3/2}\|D\| ns \rangle^2 }
{\delta E_{np_{3/2}}^2-\omega^2}+ A^{ns} ,
\label{eq2}
\end{eqnarray}
where $A^{ns}$ is independent on the value of $\omega$. All values in Eq.(\ref{eq2}) are in atomic units. 
The values of the constants $A^{ns}$ and their uncertainties 
are given in Table~\ref{tab2}. The values of $A^{ns}$ are calculated as the average values of the
totals of the $\alpha_{core}$, $\alpha_{vc}$, $\alpha_{tail}$, and 
$\alpha^{ns}_v$ terms with $n^{\prime} = n+1, n+2, n+3$ calculated in the range from 800~nm to 1600~nm.
The values of the 
$\langle np\|D\| ns \rangle$ reduced electric-dipole matrix elements
 and the corresponding energy differences $\delta E$ are also given. The electric-dipole 
matrix elements for Li are experimental values from Ref.~\cite{ADNDT}, 
all other matrix elements are experimental values compiled in \cite{relsd}.
 The energy differences are taken from Refs.~\cite{NIST, NIST1}. 
 The values of the ac polarizabilities obtained using  Eq.(\ref{eq2}) differ from our results in Table~\ref{tab1}
 in the relevant wavelength range by less than 0.1\% which is significantly below the 
 accuracy of the values themselves. We note that the formula (\ref{eq2})
 is not applicable for the lower wavelengths  owing to the importance of the other $n^{\prime}p-ns$ resonances.
 For example,  the 420~nm wavelength corresponds to $6p-5s$ transition in Rb; therefore, the ground state 
 polarizability at this frequency will be dominated by the $n^{\prime}=6$ term.  
 
 \section{The frequency-matching criteria for mixed species}

 We have also conducted a search of the wavelength values at which two species of
optically-trapped alkali atoms ($^6$Li, $^7$Li, $^{23}$Na, $^{40}$K,$^{41}$K, $^{87}$Rb, $^{133}$Cs, and $^{210}$Fr)
 would have the same frequencies of
oscillation in a common optical trap, i.e. $\lambda$ such that
\begin{equation}
s\equiv \sqrt{\frac{\alpha_1(\lambda)}{m_1}\frac{m_2}{\alpha_2(\lambda)}}\approx 1,
\label{match}
\end{equation}

\begin{figure} 
 \hbox{\centerline{\epsfxsize=3.3in\epsffile{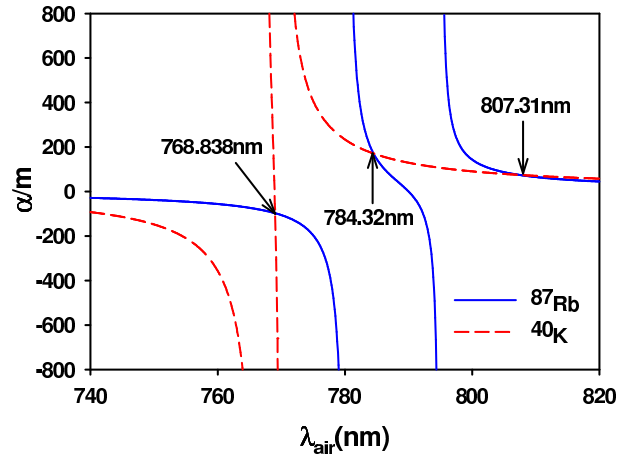}}}
   \caption{The ratios $\alpha(\lambda)/m$ of frequency-dependent polarizabilities  to the atomic weights for $^{40}$K and $^{87}$Rb.}  
    \label{fig2}     
   \end{figure} 
\noindent where $m_i$ is the atomic weight  \cite{mass}. The wavelengths of those selected matches and their uncertainties are listed in 
Table~\ref{tab3}. We have omitted all matches at wavelengths below 500~nm. While we had found numerous 
matches in the range from 300~nm to 500~nm they are difficult to
 place accurately since the accuracy of our 
calculation is lower in this region. The width of the matches in the $300 - 500$~nm region is generally
very narrow and the corresponding values of the polarizabilities are small, making them of limited experimental use.
 We define the width of the 
match as the wavelength range where the parameter $s$ given by Eq.~{(\ref{match})} is between 0.9 and 1.1.
The width of the match is very 
narrow if the corresponding wavelength is in the immediate vicinity of the  resonance.
We have also omitted most matches 
at higher wavelengths from Table~\ref{tab3} that are close to the resonance. 
Note that we listed the values of the wavelengths in the air. While the difference between 
air and vacuum wavelengths is small in the considered region, $0.15-0.4$~nm, it is sufficient to 
cause changes in the frequency-dependent polarizabilities (at some wavelengths) which exceed the uncertainties of the calculations. 
We used the following formula to compute the wavelength in air from the vacuum wavelength\cite{refractive}
\begin{eqnarray}
\lambda_{\rm air}&=&\lambda_{v}/(1+10^{-8}(8060.51+\frac{2480990}{132.274-\sigma ^2\times 10^6} \nonumber\\
&+&\frac{17455.7}{39.32957-\sigma ^2\times 10^6})).
\end{eqnarray}
The wavenumber $\sigma$ is here to be expressed in reciprocal nm.
 We also list the values of the ac polarizabilities and their uncertainties in Table~\ref{tab3}.

 The uncertainty in the wavelength of the match results from the uncertainties in the values of the frequency-dependent polarizabilities 
$\alpha_1(\lambda)$ and $\alpha_2(\lambda)$
 given in Table~\ref{tab3}. The uncertainties in the 
 values of $\alpha(\lambda)$ are found to be independent of $\lambda$
  within the range of the match to a very good precision but are generally 
  different for $\alpha_1$ and $\alpha_2$. We calculate the 
  resulting uncertainties $\delta \lambda_1$ and $\delta \lambda_2$ due to uncertainties in the values of $\alpha_1$ and  $\alpha_2$ separately and 
add the results in quadrature to obtain the final uncertainty $\delta \lambda$.
  To evaluate $\delta \lambda_1$, we calculate the $\alpha^{\pm}_1(\omega)/m_1 =  [\alpha_1(\lambda)\pm \delta \alpha_1] /m_1$ and
   then determine the wavelengths $\lambda^{\pm}$ at which $\alpha^{\pm}_1(\lambda)/m_1=\alpha_2(\lambda)/m_2$. The maximum of the differences
   between the initial match wavelength and $\lambda^{\pm}$
   is taken to be the uncertainty $\delta \lambda_1$. The evaluation of 
 $\delta \lambda_2$ is done in the same way.
 
 We illustrate several of the matches 
 given in Table~\ref{tab3} in Fig.~\ref{fig1}
 and Fig.~\ref{fig2}. The ratios $\alpha(\lambda)/m$ of frequency-dependent polarizabilities  
 to the atomic weights   for Li and Na
  in the wavelength range 
 500-800~nm are plotted in Fig.~\ref{fig1}. The corresponding matches are 
 shown by the arrows and the corresponding wavelengths are given on the plot. 
The  values $\frac{\alpha}{m}$ for $^6$Li and $^{23}$Na match at three wavelengths
in the wavelength range plotted on Fig.~\ref{fig1}. Two of the matches, at 589.397~nm
and 670.79~nm, are very close to the resonances and, therefore, are not listed
in Table~\ref{tab3} as discussed before. The match areas for those cases are extremely narrow
and the polarizability of the corresponding alkali is very sensitive to very small 
changes in the wavelength. 
 Fig.~\ref{fig2} shows similar plot to illustrate the wavelengths
 at which $^{40}$K and $^{87}$Rb have the same frequencies of
oscillation in a common optical trap in the range 740-820~nm.

 \section{Conclusions}

 In summary, we conducted a systematic study of the ground state frequency-dependent  polarizabilities
 of the alkali atoms from ultraviolet
through infrared spectral regions. The values of the ac polarizabilities and their
uncertainties are calculated for a number of wavelengths, including the  wavelengths of
common infrared lasers. A combination of 
high-precision measurements of the ratios of frequency-dependent polarizabilities of different species
could provide excellent tests of the current experimental and theoretical values of the electric-dipole matrix elements 
 in alkali-metal atoms. 
 We provide formulas and the necessary parameters  for the accurate 
calculation of the ac polarizabilities for all alkali-metal atoms at wavelengths above 800~nm.
Finally, we list selected wavelengths at which two species of
alkali atoms would have the same oscillation frequencies in a common optical trap.  \\

  \noindent  This work was performed under the sponsorship of the U.S.\ Department of Commerce, National 
   Institute of Standards and Technology. 


\begin{thebibliography}{20}
\expandafter\ifx\csname natexlab\endcsname\relax\def\natexlab#1{#1}\fi
\expandafter\ifx\csname bibnamefont\endcsname\relax
  \def\bibnamefont#1{#1}\fi
\expandafter\ifx\csname bibfnamefont\endcsname\relax
  \def\bibfnamefont#1{#1}\fi
\expandafter\ifx\csname citenamefont\endcsname\relax
  \def\citenamefont#1{#1}\fi
\expandafter\ifx\csname url\endcsname\relax
  \def\url#1{\texttt{#1}}\fi
\expandafter\ifx\csname urlprefix\endcsname\relax\def\urlprefix{URL }\fi
\providecommand{\bibinfo}[2]{#2}
\providecommand{\eprint}[2][]{\url{#2}}

\bibitem[{\citenamefont{Lundblad et~al.}(2004)\citenamefont{Lundblad, Aveline,
  Thompson, Kohel, Ramirez-Serrano, Klipstein, Enzer, Yu, and
  Maleki}}]{Lundblad}
\bibinfo{author}{\bibfnamefont{N.}~\bibnamefont{Lundblad}},
  \bibinfo{author}{\bibfnamefont{D.~C.} \bibnamefont{Aveline}},
  \bibinfo{author}{\bibfnamefont{R.~J.} \bibnamefont{Thompson}},
  \bibinfo{author}{\bibfnamefont{J.~M.} \bibnamefont{Kohel}},
  \bibinfo{author}{\bibfnamefont{J.}~\bibnamefont{Ramirez-Serrano}},
  \bibinfo{author}{\bibfnamefont{W.~M.} \bibnamefont{Klipstein}},
  \bibinfo{author}{\bibfnamefont{D.~G.} \bibnamefont{Enzer}},
  \bibinfo{author}{\bibfnamefont{N.}~\bibnamefont{Yu}}, \bibnamefont{and}
  \bibinfo{author}{\bibfnamefont{L.}~\bibnamefont{Maleki}},
  \bibinfo{journal}{J.\ Opt.\ Soc.\ Am.\ B} \textbf{\bibinfo{volume}{21}},
  \bibinfo{pages}{3} (\bibinfo{year}{2004}).

\bibitem[{\citenamefont{Schreck et~al.}(2001)\citenamefont{Schreck, Khaykovich,
  Corwin, Ferrari, Bourdel, Cubizolles, and Salomon}}]{schreck:080403}
\bibinfo{author}{\bibfnamefont{F.}~\bibnamefont{Schreck}},
  \bibinfo{author}{\bibfnamefont{L.}~\bibnamefont{Khaykovich}},
  \bibinfo{author}{\bibfnamefont{K.~L.} \bibnamefont{Corwin}},
  \bibinfo{author}{\bibfnamefont{G.}~\bibnamefont{Ferrari}},
  \bibinfo{author}{\bibfnamefont{T.}~\bibnamefont{Bourdel}},
  \bibinfo{author}{\bibfnamefont{J.}~\bibnamefont{Cubizolles}},
  \bibnamefont{and} \bibinfo{author}{\bibfnamefont{C.}~\bibnamefont{Salomon}},
  \bibinfo{journal}{Phys.\ Rev.\ Lett.} \textbf{\bibinfo{volume}{87}},
  \bibinfo{eid}{080403} (\bibinfo{year}{2001}).

\bibitem[{\citenamefont{Hadzibabic et~al.}(2002)\citenamefont{Hadzibabic, Stan,
  Dieckmann, Gupta, Zwierlein, Gorlitz, and Ketterle}}]{hadzibabic:160401}
\bibinfo{author}{\bibfnamefont{Z.}~\bibnamefont{Hadzibabic}},
  \bibinfo{author}{\bibfnamefont{C.~A.} \bibnamefont{Stan}},
  \bibinfo{author}{\bibfnamefont{K.}~\bibnamefont{Dieckmann}},
  \bibinfo{author}{\bibfnamefont{S.}~\bibnamefont{Gupta}},
  \bibinfo{author}{\bibfnamefont{M.~W.} \bibnamefont{Zwierlein}},
  \bibinfo{author}{\bibfnamefont{A.}~\bibnamefont{Gorlitz}}, \bibnamefont{and}
  \bibinfo{author}{\bibfnamefont{W.}~\bibnamefont{Ketterle}},
  \bibinfo{journal}{Phys.\ Rev.\ Lett.} \textbf{\bibinfo{volume}{88}},
  \bibinfo{eid}{160401} (\bibinfo{year}{2002}).

\bibitem[{\citenamefont{Modugno et~al.}(2002)\citenamefont{Modugno, Roati,
  Riboli, Ferlaino, Brecha, and Inguscio}}]{Modugno}
\bibinfo{author}{\bibfnamefont{G.}~\bibnamefont{Modugno}},
  \bibinfo{author}{\bibfnamefont{G.}~\bibnamefont{Roati}},
  \bibinfo{author}{\bibfnamefont{F.}~\bibnamefont{Riboli}},
  \bibinfo{author}{\bibfnamefont{F.}~\bibnamefont{Ferlaino}},
  \bibinfo{author}{\bibfnamefont{R.~J.} \bibnamefont{Brecha}},
  \bibnamefont{and} \bibinfo{author}{\bibfnamefont{M.}~\bibnamefont{Inguscio}},
  \bibinfo{journal}{Science} \textbf{\bibinfo{volume}{297}},
  \bibinfo{pages}{2240} (\bibinfo{year}{2002}).

\bibitem[{\citenamefont{Stan et~al.}(2004)\citenamefont{Stan, Zwierlein,
  Schunck, Raupach, and Ketterle}}]{Stan}
\bibinfo{author}{\bibfnamefont{C.~A.} \bibnamefont{Stan}},
  \bibinfo{author}{\bibfnamefont{M.~W.} \bibnamefont{Zwierlein}},
  \bibinfo{author}{\bibfnamefont{C.~H.} \bibnamefont{Schunck}},
  \bibinfo{author}{\bibfnamefont{S.~M.~F.} \bibnamefont{Raupach}},
  \bibnamefont{and} \bibinfo{author}{\bibfnamefont{W.}~\bibnamefont{Ketterle}}
  (\bibinfo{year}{2004}), \bibinfo{note}{cond-mat/0406129}.

\bibitem[{\citenamefont{Inouye et~al.}(2004)\citenamefont{Inouye, Goldwin,
  Olsen, Ticknor, Bohn, and Jin}}]{Inouye}
\bibinfo{author}{\bibfnamefont{S.}~\bibnamefont{Inouye}},
  \bibinfo{author}{\bibfnamefont{J.}~\bibnamefont{Goldwin}},
  \bibinfo{author}{\bibfnamefont{M.~L.} \bibnamefont{Olsen}},
  \bibinfo{author}{\bibfnamefont{C.}~\bibnamefont{Ticknor}},
  \bibinfo{author}{\bibfnamefont{J.~L.} \bibnamefont{Bohn}}, \bibnamefont{and}
  \bibinfo{author}{\bibfnamefont{D.~S.} \bibnamefont{Jin}}
  (\bibinfo{year}{2004}), \bibinfo{note}{cond-mat/0406208}.

\bibitem[{\citenamefont{Morales et~al.}(2005)\citenamefont{Morales, Nygaard,
  Williams, and Clark}}]{Miguel}
\bibinfo{author}{\bibfnamefont{M.~A.} \bibnamefont{Morales}},
  \bibinfo{author}{\bibfnamefont{N.}~\bibnamefont{Nygaard}},
  \bibinfo{author}{\bibfnamefont{J.~E.} \bibnamefont{Williams}},
  \bibnamefont{and} \bibinfo{author}{\bibfnamefont{C.~W.} \bibnamefont{Clark}},
  \bibinfo{journal}{New J. Phys.} \textbf{\bibinfo{volume}{7}},
  \bibinfo{pages}{87} (\bibinfo{year}{2005}).

\bibitem[{\citenamefont{Safronova et~al.}(2003)\citenamefont{Safronova,
  Williams, and Clark}}]{gate}
\bibinfo{author}{\bibfnamefont{M.~S.} \bibnamefont{Safronova}},
  \bibinfo{author}{\bibfnamefont{C.~J.} \bibnamefont{Williams}},
  \bibnamefont{and} \bibinfo{author}{\bibfnamefont{C.~W.} \bibnamefont{Clark}},
  \bibinfo{journal}{Phys.\ Rev.\ A} \textbf{\bibinfo{volume}{67}},
  \bibinfo{pages}{040303(R)} (\bibinfo{year}{2003}).

\bibitem[{\citenamefont{Safronova et~al.}(2004)\citenamefont{Safronova,
  Williams, and Clark}}]{rb}
\bibinfo{author}{\bibfnamefont{M.~S.} \bibnamefont{Safronova}},
  \bibinfo{author}{\bibfnamefont{C.~J.} \bibnamefont{Williams}},
  \bibnamefont{and} \bibinfo{author}{\bibfnamefont{C.~W.} \bibnamefont{Clark}},
  \bibinfo{journal}{Phys.\ Rev.\ A} \textbf{\bibinfo{volume}{69}},
  \bibinfo{pages}{022509} (\bibinfo{year}{2004}).

\bibitem[{\citenamefont{Johnson et~al.}(1983)\citenamefont{Johnson, Kolb, and
  Huang}}]{jkh}
\bibinfo{author}{\bibfnamefont{W.~R.} \bibnamefont{Johnson}},
  \bibinfo{author}{\bibfnamefont{D.}~\bibnamefont{Kolb}}, \bibnamefont{and}
  \bibinfo{author}{\bibfnamefont{K.-N.} \bibnamefont{Huang}},
  \bibinfo{journal}{At. Data Nucl.\ Data Tables} \textbf{\bibinfo{volume}{28}},
  \bibinfo{pages}{333} (\bibinfo{year}{1983}).

\bibitem[{\citenamefont{Safronova et~al.}(1999)\citenamefont{Safronova,
  Johnson, and Derevianko}}]{relsd}
\bibinfo{author}{\bibfnamefont{M.~S.} \bibnamefont{Safronova}},
  \bibinfo{author}{\bibfnamefont{W.~R.} \bibnamefont{Johnson}},
  \bibnamefont{and}
  \bibinfo{author}{\bibfnamefont{A.}~\bibnamefont{Derevianko}},
  \bibinfo{journal}{Phys.\ Rev.\ A} \textbf{\bibinfo{volume}{60}},
  \bibinfo{pages}{4476} (\bibinfo{year}{1999}).

\bibitem[{\citenamefont{Blundell et~al.}(1991)\citenamefont{Blundell, Johnson,
  and Sapirstein}}]{cs}
\bibinfo{author}{\bibfnamefont{S.~A.} \bibnamefont{Blundell}},
  \bibinfo{author}{\bibfnamefont{W.~R.} \bibnamefont{Johnson}},
  \bibnamefont{and}
  \bibinfo{author}{\bibfnamefont{J.}~\bibnamefont{Sapirstein}},
  \bibinfo{journal}{Phys.\ Rev.\ A} \textbf{\bibinfo{volume}{43}},
  \bibinfo{pages}{3407} (\bibinfo{year}{1991}).

\bibitem[{\citenamefont{Safronova et~al.}(1998)\citenamefont{Safronova,
  Derevianko, and Johnson}}]{na}
\bibinfo{author}{\bibfnamefont{M.~S.} \bibnamefont{Safronova}},
  \bibinfo{author}{\bibfnamefont{A.}~\bibnamefont{Derevianko}},
  \bibnamefont{and} \bibinfo{author}{\bibfnamefont{W.~R.}
  \bibnamefont{Johnson}}, \bibinfo{journal}{Phys.\ Rev.\ A}
  \textbf{\bibinfo{volume}{58}}, \bibinfo{pages}{1016} (\bibinfo{year}{1998}).

\bibitem[{\citenamefont{Vasilyev et~al.}(2002)\citenamefont{Vasilyev, Savukov,
  Safronova, and Berry}}]{csus}
\bibinfo{author}{\bibfnamefont{A.~A.} \bibnamefont{Vasilyev}},
  \bibinfo{author}{\bibfnamefont{I.~M.} \bibnamefont{Savukov}},
  \bibinfo{author}{\bibfnamefont{M.~S.} \bibnamefont{Safronova}},
  \bibnamefont{and} \bibinfo{author}{\bibfnamefont{H.~G.} \bibnamefont{Berry}},
  \bibinfo{journal}{Phys.\ Rev.\ A} \textbf{\bibinfo{volume}{66}},
  \bibinfo{pages}{020101(R)} (\bibinfo{year}{2002}).

\bibitem[{\citenamefont{Moore}(1971)}]{NIST}
\bibinfo{author}{\bibfnamefont{C.~E.} \bibnamefont{Moore}},
  \emph{\bibinfo{title}{Atomic Energy Levels}}, vol.~\bibinfo{volume}{35} of
  \emph{\bibinfo{series}{Natl.\ Bur.\ Stand.\ Ref.\ Data Ser.}}
  (\bibinfo{address}{U.S.\ GPO, Washington, D.C.}, \bibinfo{year}{1971}).

\bibitem[{\citenamefont{Fuhr et~al.}()\citenamefont{Fuhr, Martin, Musgrove,
  Sugar, and Wiese}}]{NIST1}
\bibinfo{author}{\bibfnamefont{J.~R.} \bibnamefont{Fuhr}},
  \bibinfo{author}{\bibfnamefont{W.~C.} \bibnamefont{Martin}},
  \bibinfo{author}{\bibfnamefont{A.}~\bibnamefont{Musgrove}},
  \bibinfo{author}{\bibfnamefont{J.}~\bibnamefont{Sugar}}, \bibnamefont{and}
  \bibinfo{author}{\bibfnamefont{W.~L.} \bibnamefont{Wiese}},
  \bibinfo{note}{{\it NIST Atomic Spectroscopic Database}, NIST Physical
  Reference Data, National Institute of Standards and Technology. Available
  online, URL = http://physics.nist.gov/cgi-bin/AtData/main\_asd.}

\bibitem[{\citenamefont{Johnson et~al.}(1988)\citenamefont{Johnson, Blundell,
  and Sapirstein}}]{Bspline}
\bibinfo{author}{\bibfnamefont{W.~R.} \bibnamefont{Johnson}},
  \bibinfo{author}{\bibfnamefont{S.~A.} \bibnamefont{Blundell}},
  \bibnamefont{and}
  \bibinfo{author}{\bibfnamefont{J.}~\bibnamefont{Sapirstein}},
  \bibinfo{journal}{Phys.\ Rev.\ A} \textbf{\bibinfo{volume}{37}},
  \bibinfo{pages}{307} (\bibinfo{year}{1988}).

\bibitem[{\citenamefont{Johnson et~al.}(1996)\citenamefont{Johnson, Lui, and
  Sapirstein}}]{ADNDT}
\bibinfo{author}{\bibfnamefont{W.~R.} \bibnamefont{Johnson}},
  \bibinfo{author}{\bibfnamefont{Z.~W.} \bibnamefont{Lui}}, \bibnamefont{and}
  \bibinfo{author}{\bibfnamefont{J.}~\bibnamefont{Sapirstein}},
  \bibinfo{journal}{At. Data and Nucl. Data Tables}
  \textbf{\bibinfo{volume}{64}}, \bibinfo{pages}{279} (\bibinfo{year}{1996}).

\bibitem[{\citenamefont{Coursey et~al.}()\citenamefont{Coursey, Schwab, and
  Dragoset}}]{mass}
\bibinfo{author}{\bibfnamefont{J.~S.} \bibnamefont{Coursey}},
  \bibinfo{author}{\bibfnamefont{D.~J.} \bibnamefont{Schwab}},
  \bibnamefont{and} \bibinfo{author}{\bibfnamefont{R.~A.}
  \bibnamefont{Dragoset}}, \bibinfo{note}{{\it NIST Nuclear Physics Database},
  NIST Physical Reference Data, National Institute of Standards and Technology.
  Available online, URL =
  http://physics.nist.gov/PhysRefData/Compositions/index.}

\bibitem[{\citenamefont{Peck and Reeder}(1972)}]{refractive}
\bibinfo{author}{\bibfnamefont{E.}~\bibnamefont{Peck}} \bibnamefont{and}
  \bibinfo{author}{\bibfnamefont{K.}~\bibnamefont{Reeder}},
  \bibinfo{journal}{J.Opt.Soc.Am.} \textbf{\bibinfo{volume}{63}},
  \bibinfo{pages}{958} (\bibinfo{year}{1972}).

\end{thebibliography}
\end{document}